
\magnification 1200
\parindent=0pt

\null\hskip 10cm UB-ECM-PF-38-94\par
\null\hskip 10cm December 1994\par
\null\vskip 1cm

\centerline {\bf IS THERE PHYSICS IN LANDAU POLES?}\par
\null\vskip 1pt
\null\hskip 6cm Rolf Tarrach\par
\null\vskip 1pt
\null\hskip 3cm {\it Dep. d'ECM}, {\it Fac. de F\'\i sica}, {\it Univ.
de Barcelona}\par
\null\hskip 3cm {\it Postal address: Diagonal  647}, {\it 08028
Barcelona, Spain.}\par
\null\hskip 3cm {\it E-mail: ROLF at ECM.UB.ES}\par
\null\vskip 1pt
\centerline {and {\it IFAE}}\par
\null\vskip 1pt
\null\hskip 3cm PACS 11.10.Gh, 03.65.-w, 11.10.Jj\par
\null\vskip 2cm
\centerline {\bf ABSTRACT}\par
\null\vskip 1pt

\null\hskip 1cm Triviality and Landau poles are often greeted as
harbingers of new
physics at 1 TeV. After briefly reviewing the ideas behind this, a model
of singular quantum mechanics is introduced. Its
ultraviolet structure, as well as some features of its vacuum, related
to triviality, very much parallel $\lambda\phi^4$. The model is
solvable,
exactly and perturbatively, in any dimension. From its analysis we learn
that Landau poles do not appear in any exactly computed observable, but
only in truncated perturbation theory, when perturbation theory is
performed with the wrong sign coupling. If these findings apply to the
standard model no new physics at 1 TeV should be expected but only
challenges for theorists.
\vfill
\eject

\underbar {I. Introduction}\par

\null\hskip 1cm Our first model of weak interactions was given by the
four fermion Fermi interaction [1] with a dimensionful coupling $G_F$.
As it was not perturbatively renormalizable, one could not make much
quantum field-theoretical sense out of it. This was in sharp contrast
with the situation of electromagnetic lepton interactions, for which
quantum electrodynamics (QED) was so successful. The search of a
perturbatively renormalizable theory of weak interactions, together with
symmetry considerations, led to the Weinberg-Salam (WS) theory of
electroweak interactions [2], which subsumed QED. The Fermi model was at
once understood (modulus confinement) as part of the low energy
effective theory which remains once one integrates out the massive gauge
fields, $W^\pm$ and $Z^o$ which got their mass form the Higgs mechanism.
This leads to $G_F={1\over{\sqrt{2}v^2}}\sim M^{-2}_W$, $v$ being the
vacuum expectation value of the scalar field, and a range of validity of
the Fermi model given by $E<M_W$.\par
\null\hskip 1cm The WS theory works remarkably well up to the maximum
present day energies, but it contains the seeds of difficulties at
energies approaching the TeV if the Higgs particle is not found by then.
These difficulties go under the heading
of Landau poles [3], of triviality [4], or of growing coupling, or
equivalently, breakdown of perturbation theory [5]. These are of course
related phenomena [6] due to the scalar or Higgs sector of the WS
theory.
The central question is: are they telling us something about new physics
at $E \sim$ 1 TeV or are they indicating that perturbation theory and
some of its hidden remnants in non-perturbative studies [7], like the
relation between the Higgs coupling and mass and the vacuum expectation
value$$\lambda={M^2_H\over2 v^2}\eqno (1)$$ should be given
up or more properly
understood? Certainly the first scenario is much more exciting, and
probably theoretically easier, once the new physics is known. The second
scenario only promises theoretical difficulties in explaining smooth
experimental data at  1 TeV. It is understandably less popular, and
considered by some
somewhat heterodox. Contributing to its understanding is the aim of this
work.\par
\null\hskip 1cm Let us go a bit more into these two alternatives. Recall
that new physics has, up to now, never come at the hand of Landau poles
or triviality. Electron QED, obtained from lepton QED by integrating out
muons and taus is an effective theory which has a perturbatively
renormalizable truncation, valid for $E < m_\mu$. The new physics comes
at $E \simeq m_\mu$, zillions of orders of magnitud below its Landau
pole, which is beyond the Planck mass. Five flavour quantum
chromodynamics (QCD), obtained by integrating out the top quark, is an
effective theory with a pertubatively renormalizable truncation, valid
for $E < m_t$. The new physics comes at $E \simeq m_t$, and QCD has no
Landau pole. New physics often does not mean a new particle, but a new
vacuum. Again QCD provides examples at the chiral symmetry breaking
scale, $\Lambda_\chi$, and at the confinement scale, $\Lambda_{QCD}$.
These thresholds, with their effective theories below them, for $E <
\Lambda_{QCD}$ [8] and for $\Lambda_{QCD} < E < \Lambda_\chi$ [9], are
totally unrelated to any Landau pole. Thus, these effective theories,
effective \`a la Weinberg [10], do not provide justification for
expecting Landau pole suggested new physics at $E \sim$ 1 TeV. In fact
the very WS theory
is expected to change its vacuum at these energies which correspond
to the electroweak symmetry breaking scale, $\Lambda_{EW}$. But this is
unrelated to Landau poles too.\par
\null\hskip 1cm What about the effective theories \`a la Wilson [11],
the ones the renormalization group produces? They are related to the
previous ones, but different [12]. They are valid for $E < a^{-1}$, $a$
being the distance up to which one has integrated out the short distance
degrees of freedom. The distance $a$ is arbitrary. There is no
qualitative
new physics at $E \sim a^{-1}$, just more of the old stuff. They often
have truncations which are perturbatively renormalizable, as when one
only keeps relevant and marginal terms around a gaussian fixed point.
The couplings depend on $a$. What happens if the theory is trivial?
Could then triviality lead to new physics?\par
\null\hskip 1cm Let us shortly recall in the setting of renormalization
group improved perturbation theory the standard logic in favour of new
physics. Recalling that there does not exist a strong coupling lattice
Higgs model which can be regarded an effective continuum theory at low
energies [7] our setting should be qualitatively sound. For $\lambda
\phi^4$ the bare coupling is given to one loop by
$$\lambda_o(a)={\lambda(M)\over1+{3\lambda(M)\over 16\pi^{2}}ln (M
a)}\eqno (2)$$ so that at
$$a_L=M^{-1} exp
  [-{16\pi^2\over3\lambda(M)}]\eqno (3)$$ the bare
coupling diverges. This is the Landau pole, which, for $\lambda (M)>0$,
lies on the way towards the continuum limit, $a \rightarrow 0$. One
could insist in keeping
$\lambda_o(a)>0$ bounded as $a \rightarrow 0$, say $\lambda_o(a)<1$, so
that perturbation theory makes sense. To see how this works, invert (2),
$$\lambda(M)={\lambda_o(a)\over1-{3\lambda_o(a)\over 16\pi^{2}}ln
(Ma)}\eqno (4)$$ Now, $\lambda_o(a)$ is an adimensional
function of $a\Lambda$, $\Lambda$ being the physical scale which
quantifies the strength of the interaction, as $\Lambda_{QCD}$ does for
QCD. Then one can take the $a \rightarrow 0$ limit keeping $a\Lambda$
fixed,
which thus keeps $\lambda_o(a)$ fixed, and bounded. The result is

$$\displaystyle \lim_{{a\rightarrow
0}\atop{a\Lambda:\, {\rm fixed}}}{\lambda(M)}= 0 \eqno (5)$$


We have avoided the Landau pole, but no interaction is left. This is
known as the Moscow zero. The theory is trivial.\par
\null\hskip 1cm In the Dashen - Neuberger approximation [13], where the
gauge and Yukawa couplings are considered a small perturbation, the
above results
for O(4) scalar fields apply to the WS theory. Then (1) and (5) are only
compatible if one keeps $a \not= 0$, and the WS theory is an effective
theory \`a la Wilson without non-trivial continuum limit as necessarily
$a
> a_H, a_H$ being given by (4) (slightly modified for O(4)) and (1) with
$\lambda(M)=\lambda$, $M_H=M$ and $\lambda_o$ fixed, and bounded. The
minimal spacing $a_H$ depends
on $M_H$. However, as long as $$M^{-1}_ H > a_H \eqno (6)$$ it has
continuum
physics. As $M_H$ increases the left hand side of (6) decreases and
approaches
the right hand side. At $M_H \sim$1 TeV (6) ceases to be valid
and continuum physics vanishes. Something must happen if the Higgs is
not found by then: new physics. But even if the Higgs is found
before, by the same argument new physics is expected for energies
$E\sim a_H^{-1}$.\par
\null\hskip 1cm This seems a tight scenario, but never before new
physics came this way. So alternatives should not be dismissed
beforehand. There are alternatives. They are based on giving up (1),
which is certainly incompatible with the continuum limit.
Recall that triviality does not mean to have a symmetric vacuum, one can
have triviality and spontaneous symmetry breaking at the same time. One
point functions can be non-vanishing, only three and more point proper
Green functions have to vanish [14]. This allows to take the contiuum
limit, $a \rightarrow 0$, with $\lambda = 0$ but still $v \not= 0$ and
$M_W
\not= 0$. Concrete ways of how to implement this have been worked out
[15]. Even some lattice computations hinting in this direction have been
published [16]. The Higgs sector would be very weakly interacting and no
new physics would be expected. The existence of the Landau pole would be
unrelated to new physics.\par
\null\hskip 1cm This alternative scenario leads to important new
questions:
Is there any physics then in the Landau pole? What is the meaning of
perturbation
theory if the theory is trivial? It is these questions we want to
address here.\par
\null\hskip 1cm It is basically impossible to solve these issues for our
3 + 1 dimensional quantum field theories (QFT), where we cannot do
much more than to compute a few orders of perturbation theory. We will
thus study a model of singular quantum mechanics [17], where
everything can be worked out exactly, and which has an ultraviolet (UV)
structure which matches very precisely the one of $\lambda\phi^4$.
The picture which emerges is the following: the Landau pole is a feature
which only appears in non-observable magnitudes, or in observable ones
computed approximately, as one always does in perturbation theory. In
the exact expressions for observables there are no Landau poles. The
only meaning of the Landau pole is that one started with the wrong sign
coupling in doing perturbation theory, so that in converging to the
exact result the coupling has to change sign and does so at the Landau
pole. The Landau pole would then only be a feature of truncated
perturbation theory and a nightmare for theorists, but experimentalists
would see nothing of it.\par
\null\hskip 1cm In carrying over our findings, which are exact and
rigorous, to the WS theory there are two steps to be taken. First to go
from singular quantum mechanics to $\lambda\phi^4$. This means
going from a theory with only coupling constant renormalization to one
with several independent renormalizations. It also means going from a
perturbative expansion with a finite radius of convergence to one which
is asymptotic.
It finally means going from a theory without
non-perturbative contributions to one which will have non-perturbative
contributions, like the ones dues to instantons [18] and maybe
renormalons [19].
The other differences are more likely to be irrelevant to the problem at
hand. Second to go from $\lambda\phi^4$ to WS. If our findings
should be relevant to the WS theory we have to assume that also in
presence of gauge symmetries triviality of the scalar sector is
compatible with the Higgs mechanism in the continuum limit.\par
\null\hskip 1cm Our model are contact interactions in d = 2 quantum
mechanics (QM), $V = \lambda\delta^{(2)} ({\vec r})$. We will
first
give six reasons, some of them related, of why we believe it to be a
faithful model for the UV structure and high energy behaviour, and even
for the vacuum, of $\lambda\phi^4$ in D = 3+1.\par
\null\hskip 1cm 1. Both interactions are classically scale
invariant.\par
\null\hskip 1cm 2. Both require regularization (or alternatively
selfadjoint extensions and differential renormalization [20]
respectively) and show dimensional transmutation [21]. The role played
by the infinite number of degrees of freedom of $\lambda\phi^4$ is
played by the singular character of $\lambda \delta^{(2)}(
{\vec r})$ in quantum mechanics. Regularization puts them on a somewhat
equivalent footing.\par
\null\hskip 1cm 3. Both are perturbatively renormalizable.\par
\null\hskip 1cm 4. Both are very likely trivial for $\lambda > 0$.\par
\null\hskip 1cm 5. Both very likely collapse for $\lambda < 0$, but are
asymptotically free [22].\par
\null\hskip 1cm 6. Both are given at their critical dimensions in the
random
walk picture. In the random walk representation of $\lambda\phi^4$
the interaction is given by the crossing of two random paths [23]. As
the Hausdorff dimension of the random path is $d_H =
2$, $D = 3+1 = 2d_H$ is the critical dimension. Thus for $D > 2d_H$ the
theory is trivial and for $D < 2d_H$ it is interacting. For $\lambda
\delta^{(2)}({\vec r)}$ the interaction is given by the
crossing
of the origin by the path in the path integral. Again $d = d_H$ is the
critical dimension.\par
\null\hskip 1cm The triviality issue of $\lambda\phi^4$ has been
studied by taking its non-relativistic limit, which leads to
$V={\lambda\over m}\delta^{(3)}({\vec r)}$ [24]. We feel that
the non-relativistic limit is not adequate for studying an UV issue as
triviality. Most of the above reasons do not hold. On the other hand
$\lambda \delta^{(2)}({\vec r})$, $\lambda < 0$, has been used
as a model for asymptotic freedom [25]. Here we are interested mainly in
$\lambda > 0$, where Landau poles appear, because $\lambda\phi^4$
is just a sick theory for $\lambda < 0$ when taken as it stands
[14].\par
\null\hskip 1cm To buttress the analogy of $\lambda
\delta^{(2)}({\vec r})$ and $\lambda\phi^4_{3+1}$ we will
study $\lambda \delta^{(d)}({\vec r})$ for d = 1, 2 and 3 [25,
26]
and see how its UV structure matches the one of $\lambda\phi^4_ D$
with $D < 3+1$, D = 3+1 and $D > 3+1$ respectively. Our units are $\hbar
= 1$, $2m = 1$.\par\null\vskip 1pt
\underbar {II. d = 1}\par
\null\hskip 1cm The potential $V(x)=\lambda\delta(x)$ is solved in many
textbooks [27]. Although singular it does not require regularization.
For $\lambda < 0$ there is one bound state of energy $$E_o = -
{\lambda^2\over 4}\eqno (7)$$ For $E= k^2\geq
0$ the scattering amplitude $f(k)$ is defined as
$$\Psi_k(x)_
{\mid x\mid \rightarrow\infty}\sim {e^{ikx}+e^{i(k\mid
x\mid+{\pi\over 2})}(f_+(k)\theta (x)+f_-(k)\theta (-x))}\eqno (8)$$
For the Dirac delta potential $f_+=f_-\equiv f$ reads, for any
$\lambda$, $$f(k)\equiv{e^{2i\delta_o}-1\over 2i}={-\lambda\over
2k+i\lambda}\eqno (9)$$ which leads to the even phase shift (the odd
phase shift vanishes) $$tg \delta_o(k)=-{\lambda\over 2k}\eqno
(10)$$\null\hskip
1cm At high energies there is no scattering, $$\displaystyle \lim_{
k>>\lambda}\delta_o(k)=0 \eqno (11)$$ The connection between
high and low energies is correctly given by Levinson's theorem
$${1\over 2}+{1\over \pi}[\delta_o(0)-\delta_0(\infty)]=1,\null\hskip
1cm \lambda<0\eqno (12)$$ and counts the number of bound
states.\par
\null\hskip 1cm Perturbation theory is best built upon the Lippmann -
Schwinger equation, which for any $d$ reads $$\Psi(
{\vec r})=\Psi_o({\vec r})-\int d^d r' G_{k+}(
{\vec r}-{\vec r\prime}) V ({\vec r\prime})\Psi
({\vec r\prime})\eqno
(13)$$ where $\Psi_o$ is a solution of the free Schr\" odinger equation
and $G_{k+}$ the free progagator. For d=1
$$G_{k+}(x)=i{e^{ik\mid x\mid}\over 2k}\eqno (14)$$ is the Green's
function with the adequate large $\mid x\mid$ behaviour. From (8), (13)
and (14) one obtains in perturbation theory $$f(k)= -{\lambda\over 2k}+
i{\lambda^2\over 4k^2}+{\lambda^3\over (2k)^3}+...\eqno (15)$$
This is a geometric series which can be summed at high energies, $2k >
\mid\lambda\mid$. The result is the exact result (9), which is thus
reproduced by analytic continuation.\par
\null\hskip 1cm The theory is effectively asymptotically free as the
expansion (15) is actually in the dimensionless function $$\hat\lambda
(k)\equiv{\lambda\over 2k}\eqno (16)$$ which vanishes for high
energies. One can always introduce a dimensionless constant, the
renormalized coupling
$$\hat\lambda_r(\mu)\equiv\hat\lambda(k=\mu)\eqno (17)$$ where
$\mu$ is an arbitrary scale. Then (9) reads
$$f(k)={-\hat\lambda_r(\mu)\over
{k\over\mu}+i\hat\lambda_r(\mu)}={-\hat\lambda(k)\over
1+i\hat\lambda(k)}\eqno (18)$$ The $\mu$- dependence in $f(k)$ is of
course fictitious.\par
\null\hskip 1cm At low energies $\hat\lambda$(k) blows up. Notice
however
that none of the observables, i.e. the scattering amplitude, in any
sense reflects this singularity. This is not so for the perturbative
result (15), which does not hold at low energies, where it diverges.\par
\null\hskip 1cm So far the two body problem. What happens with the N
body problem when N becomes large with fixed density? It is known [28]
that N bosons in a space $0\leq x_i\leq L$ interacting via attractive
two body contact interactions $(\lambda < 0)$ have a ground state energy
per particle which goes as

$$ \displaystyle \lim_{ { N\rightarrow \infty } \atop { {{N} \over {L}}
:\,{\rm fixed} } }{ {E_0 (N)}\over{N}} \sim -N \eqno (19)$$


 This implies that no vacuum exists for
$\lambda<0$. We call
vacuum in QM the state of lowest energy per particle for any N at zero
particle density or for any particle density as $N\rightarrow \infty$.
It is this
analysis in QM which corresponds to studying the existence of the vacuum
in QFT for $\lambda <0$.\par
\null\hskip 1cm $\lambda \phi^4_ {2+1}$ is a very similar QFT. It
is perturbatively superrenormalizable and only the vacuum and the mass
require renormalization up to a finite order. As the coupling constant
is not renormalized this corresponds to no renormalization in QM. The
coupling has dimensions and is asymptotically free in very much the same
sense as above. The theory is not trivial, it interacts for $\lambda>0$.
For $\lambda<0$ no ground state exists, but perturbation theory goes
through.\par
\null\hskip 1cm One might think that at one point the analogy breaks
down for $\lambda<0$, as QM allows a bound state but QFT has no ground
state, and therefore does not make sense. This is because in QM, being a
theory which does not allow creation and annihilation of particles, a
collapse for $N\rightarrow \infty$ does not affect the study of the two
body problem, and the existence of its possible bound states. In QFT,
not having a ground state spoils any sensible physical content of the
theory. But it is precisely this feature of QM, and its solvability,
which allows the study we are perfoming, and leads to the understanding
we will gain along this work.\par\null\vskip 1pt
\underbar {III. d=2} [29]\par
\null\hskip 1cm We know that $V({\vec r})=\lambda
\delta^{(2)}({\vec r})$ requires regularization. We will do so by
writing $$V_R(r)={\lambda\over 2\pi R}\delta (r-R),\null\hskip 1cm
R>0\eqno
(20)$$ which for $R\rightarrow 0$ reproduces $V({\vec r})$. Finite
results for physical magnitudes as $R\rightarrow 0$ are obtained only if
$$\lambda (R)={2\pi\over ln {R\over R_o}}<0,\null\hskip 1cm R<R_o\eqno
(21)$$
where subleading terms as $R\rightarrow 0$ are irrelevant and have been
dropped and where $R_o$ is a length which characterizes the strength
of the interaction (as $\Lambda_{QCD}$ does for strong interactions).
Notice
that $\lambda (R)$ is negative. It is called the bare coupling. It leads
to a bound state of energy $$E_o = -{4\over
R^2_o}e^{-2\gamma}\eqno (22)$$ $\gamma$ being Euler's
constant. $\lambda (R)$ has been traded for $R_o$; this is dimensional
transmutation.\par
\null\hskip 1cm One might want to reintroduce a coupling, as a
renormalized coupling, $$\lambda_r(\mu)\equiv {\lambda (R)\over
1-{\lambda
(R)\over 2\pi}[ln{\mu R\over 2}+\gamma]}={-2\pi\over ln{\mu R_o\over
2}+\gamma}\eqno (23)$$ in terms of which (22) reads.\par
$$E_o=-\mu^2 exp{4\pi\over \lambda_r(\mu)}\eqno (24)$$ Notice
that $E_o$ does not depend en $\mu$, which is arbitrary and physically
irrelevant, but on $\lambda_r(\mu)$. The only relevance of $\mu$ comes
from it having dimensions. As $$\lambda_r(\mu={2\over
R}e^{-\gamma})=\lambda (R)\eqno (25)$$ bare and renormalized
couplings are basically the same, with one caveat: the regulator R is
introduced for small distances, $R<R_o$, while $\mu$ can take any
value.\par
\null\hskip 1cm For scattering, $E=k^2>0$, again only $\lambda (R)$ as
given by (21) leads to finite results as $R\rightarrow 0$. One obtains
for the scattering amplitude, defined as $$\Psi ({\vec
r})_{r\rightarrow\infty}\null\hskip 1pt\sim\null\hskip 1pt
e^{i {\vec k}\cdot{\vec r}}+{1\over \sqrt{r}}e^{i(kr+{\pi\over
4})}f(k,\theta)\eqno (26)$$
the following result $$f(k,\theta)=\sqrt {{\pi\over 2k}}{1\over
ln{kR_o\over2}+\gamma - {i\over 2}\pi}={-1\over
2\sqrt{k\pi}}{\lambda_r(\mu)\over 1+\lambda_r(\mu)[{i\over 4}-{1\over
2\pi}ln{k\over \mu}]}\eqno (27)$$ It is convenient to
introduce a running coupling constant, a function in fact,
$$\lambda_r(k)=\lambda_r(\mu=k)\eqno (28)$$ so that (27) reads
$$f(k,\theta)={-1\over 2\sqrt {k\pi}}{\lambda_r (k)\over 1+{i\over
4}\lambda_r (k)}\eqno (29)$$ which displays the $\mu-$
independence of the scattering amplitude as well as the absence of
logarithms of $k$ when the running coupling is used. Only s- waves
scatter.\par
\null\hskip 1cm From the partial wave expansion
$$f(k,\theta)={-i\over \sqrt{2\pi k}}\sum_{l=-\infty}^\infty
(e^{zi\delta_l}-1)e^{il\theta}\eqno (30)$$ one obtains
$$tg\delta_o(k)={\pi\over 2(ln{kR_o\over 2}+\gamma)}={\pi\over
2(ln{k\over \mu}-{2\pi\over \lambda_r(\mu)})}=-{\lambda_r(k)\over
4}\eqno (31)$$
\null\hskip 1cm There is no scattering at high energies,
$$\displaystyle\lim_{kR_o\rightarrow \infty}\delta_o(k)=0\eqno (32)$$
and Levinson's theorem
holds, $${1\over \pi}[\delta_o(0)-\delta_o(\infty)]=1\eqno
(33)$$
\null\hskip 1cm The result (32) reflects asymptotic freedom,
$$\displaystyle\lim_{R<<R_o}\lambda (R)=0$$
$$\displaystyle\lim_{kR_o\rightarrow \infty}\lambda_r(k)=0\eqno
(34)$$ Notice the existence of a harmless Landau pole,
$$\lambda(R_L=R_o)=\infty$$ $$\lambda_r(\mu_L=k_L={2\over
R_o}e^{-\gamma})=\infty\eqno (35)$$ Since $R$ can always be taken
smaller than $R_o$ and $\mu$ does not show up in observables, the Landau
pole is innocuous. One
can indeed go all the way to $k=0$ in (29) or
(31) without encountering singularities in physical magnitudes. The
coupling $\lambda_r(k)$ however changes sign and becomes positive.
Notice that $$k^2_L=-E_o\eqno (36)$$ The existence of the Landau pole is
therefore
linked to the existence of a bound state, which of course only exists
for attractive interactions.\par

\null\hskip 1cm To work out perturbations we only need the propagator
$$G_{k+}({\vec r})={i\over 4}H^{(1)}_o(kr)\eqno (37)$$
where $H^{(1)}_o$ is the first Hankel function of zero order. To first
order one obtains $$f^{(1)}(k)=-{\lambda_p\over 2\sqrt {2\pi
k}}+O(R)\eqno (38)$$ where the coupling of (20) has been
rebaptized $\lambda_p$. At second order it diverges as $Rk\rightarrow
0$, $$f^{(2)}(k)={\lambda_p^2\over 2\sqrt{2\pi k}} ({i\over 4}-{1\over
2\pi}(ln{kR\over 2}+\gamma))+O(R)\eqno (39)$$ In order to get
a finite result one has to add a counterterm to the interaction:
$$V_R(r)=V_R^{(1)}(r)+V_R^{(2)}(r)={\lambda_p\over 2\pi R}(1-{\lambda_
p\over 2\pi}(ln{R\nu\over 2}+\gamma))\delta(r-R)\eqno (40)$$
where $\nu$ is an arbitrary scale. As $V_R$ should not depend on it,
$\lambda_p$ will have to, $\lambda_p(\nu)$. The first and second order
result obtained from (40), which is finite by construction and called
renormalized, is $$f_r(k)=-{\lambda_p(\nu)\over 2\sqrt{2\pi
k}}[1-\lambda_p(\nu)({i\over 4}-{1\over 2\pi}ln{k\over
\nu})]+O(\lambda^3_p)\eqno (41)$$ One can proceed to next
order. Geometric series appear both for $V_R(r)$ and $f_r(k)$. The
results
of the summations are $$V_R(r)={1\over 2\pi R}{\lambda_p(\nu)\over
1+{\lambda_p(\nu)\over 2\pi}(ln{R\nu\over
2}+\gamma)}\delta(r-R)\eqno (42)$$ and $$f_r(k)={-1\over
2\sqrt{\pi k}}{\lambda_p(\nu)\over 1+\lambda_p(\nu)({i\over 4}-{1\over
2\pi}ln{k\over \nu})}\eqno (43)$$ Both reproduce the exact
results with the identification $$\lambda_p(\nu =
\mu)=\lambda_r(\mu)\eqno (44)$$ As for d=1 the exact result is
an analytic continuation of the summed perturbative one.\par
\null\hskip 1cm Notice that perturbation theory can be performed both
for $\lambda_p>0$ and $\lambda_p<0$. And yet the interaction is
eventually always attractive. This makes the Landau pole more
significant in perturbation theory. From eq. (42) and for $\lambda_p>0$
but $\lambda_p<<1$ one encounters a singularity as $R\nu \rightarrow 0$.
And yet for the scattering amplitude, eq. (43), which is the only
predicted observable ($E_o$ is used to fix $R_o$), no singularity
appears anywhere. The Landau pole is irrelevant to physics as given by
the exact expressions. What it tells us is that if $\lambda_p>0$ but the
interaction is actually attractive, the coupling will have to change
sign and it does so at the Landau pole. The Landau pole thus only has to
be crossed in going from perturbation theory to the exact theory if one
started with a coupling of the wrong sign. As seen from eq. (36), when
it exists it implies that the true interaction is attractive. On the
contrary, if one starts from $\lambda_p<0$ one does not have to cross
the Landau pole in taking the continuum limit $R\nu \rightarrow 0$. But
one always sees
the Landau pole in $\lambda_r(k)$ at $k_L$ given by eq. (36). We have
learned, however, that the observables do not notice the Landau pole;
the exactly computed observables, of course. Any perturbative truncation
will make the observables notice the Landau pole.\par
\null\hskip 1cm What happens for the N- body problem? It is only known
that for finite N the system does not collapse, as a lower bound exists
[30]. There are strong reasons, however, to believe that as
$N\rightarrow \infty$ with fixed density, the ground state energy per
particle is unbounded from below. These reasons, based on continuity in
going from d=1 to d=3, will become clearer when we study next d=3. If we
accept these reasons, then no vacuum exists for $\lambda<0$ in QM.\par
\null\hskip 1cm $\lambda \phi^4_{3+1}$ is a QFT of analogous
features. It is perturbatively renormalizable both for $\lambda>0$ and
$\lambda<0$. It requires vacuum, mass, field and coupling constant
renormalization. In QM this means coupling constant renormalization. For
$\lambda>0$ there is a worrisome Landau pole and the theory is believed
to be trivial. For $\lambda<0$ the exact theory will not exist as such.
QM gives us a model of what could happen: the Landau pole is actually
not a problem, but it signals that although $\lambda_p>0$ eventually
$\lambda<0$. This is fine in finite N QM, but does not allow to define
the QFT, which is thus trivial, $\lambda=0$.\par\null\vskip 1pt
\underbar {IV. d=3}\par
\null\hskip 1cm We will use as regulator of $V({\vec r})=\lambda
\delta^{(3)}({\vec r})$\par
$$V_R({\vec r})={\lambda\over 4\pi R^2}\delta(r-R),\null\hskip 1cm
R>0\eqno (45)$$
Finiteness as $R\rightarrow 0$ requires (omitting irrelevant subleading
terms) $$\lambda(R)=-4\pi R(1+{R\over R_o})<0,\null\hskip 1cm R<\mid
R_o\mid\eqno
(46)$$ For $R_o>0$ there is a bound state of energy $$E_o=-{1\over
R^2_o}\eqno (47)$$ but the interaction is always attractive,
even when $R_o<0$.\par
\null\hskip 1cm The scattering amplitude, defined as $$\Psi (
{\vec r})_{r\rightarrow\infty} \sim e^{i{\vec k}\cdot {\vec r}}+{1\over
r}e^{ikr}f(k,\theta,\varphi)\eqno (48)$$
comes out to be $$f(k)={-R_o\over
1+iR_ok}\eqno
(49)$$ Again only s-waves scatter. From the partial wave expansion for
rotationally invariant potentials $$f(k,\theta)={1\over
2ik}\sum^\infty_{l=0}(2l+1)(e^{2i\delta_l}-1)P_l(cos\theta)\eqno
(50)$$ one obtains $$tg \delta_0(k)=-R_ok\eqno (51)$$
Notice that there is scattering at high energies,
$$\displaystyle\lim_{kR_o\rightarrow
\infty}\delta_o(k)\not=0 \eqno (52)$$ This is a surprising
result, as the theory looks asymptotically free,
$$\displaystyle\lim_{R\rightarrow
0}\lambda(R)=0\eqno (53)$$ As $\lambda (R)$ has dimensions
this result is not very meaningful. Let us introduce a dimensionless
coupling which weighs the UV behaviour of $V_R(r)$, $$\hat\lambda
(R)\equiv {\lambda (R)\over R}\eqno (54)$$ Then $$\displaystyle\lim_
{R<<R_o}{\hat\lambda}(R)=-4\pi \not= 0\eqno (55)$$ and there is no
asymptotic
freedom. There is interaction, even at high energies, and (52) follows.
Levinson's theorem does not hold either.\par
\null\hskip 1cm Had we taken the limit $k\rightarrow \infty$ before
$R\rightarrow 0$ the result would have been
$$\displaystyle\lim_{kR_o\rightarrow \infty}\delta_o
(k,R)=0\eqno (56)$$ The limits $k\rightarrow \infty$ and
$R\rightarrow 0$ do not commute. This is a feature which has
catastrophic consequences for three particle systems, known as Thomas
effect [31] (see ref. 32 for a more modern presentation): in d = 3 the
binding energy of the three-particle system increases beyond limit as
the range of the two-body short range interaction goes to zero. In other
words,
non-trivial contact interactions are just too strong in d = 3. Eq. (52)
hints at it for two particles, the Thomas effect shows it bluntly for
three particles. No vacuum exists. Only in the two body channel
some, somewhat weird, physics remains.\par
\null\hskip 1cm Furthermore, there is no Landau pole. This is an ominous
result. Recall that perturbation theory is blind to the sign of
$\lambda_p$, and that it was the Landau pole where the interaction
becomes attractive.
Since there is no Landau pole, and the exact interaction
is attractive, something must go wrong with perturbations. Indeed, it
does.\par
\null\hskip 1cm The propagator is
$$G_{k+}({\vec r})={1\over 4\pi r}e^{ikr}\eqno (57)$$
First order perturbation theory gives $$f^{(1)}(k)=-{\lambda_p\over
4\pi}+0(R^2)\eqno (58)$$ while at second order the result is
UV divergent, $$f^{(2)}(k)=({\lambda_p\over 4\pi})^2({1\over
R}+ik)+O(R)\eqno (59)$$ Adding a counterterm to the
interaction,
$$V_R(r)={\lambda_p\over 4\pi R^2}(1+{\lambda_p\over 4\pi R})\delta
(r-R)\eqno (60)$$ gives $$f_r(k)=-{\lambda_p\over
4\pi}+i({\lambda_p\over 4\pi})^2 k+0(\lambda^3_p)\eqno (61)$$
To third order one obtains $$f^{(3)}(k,\theta)=({\lambda_p\over
4\pi})^2({1\over R^2}+{2\over 3}k^2+{5\over 27}k^2 cos\theta)\eqno
(62)$$Notice the $\theta$-dependence! A new counterterm is added to
(60) $$V_R(r)={\lambda_p\over4\pi R^2}(1+{\lambda_p\over 4\pi
R}+({\lambda_p\over 4\pi
R})^2)\delta(r-R)\eqno (63)$$ This gives
$$f_r(k,\theta)=-{\lambda_p\over 4\pi}+i({\lambda_p\over 4\pi})^2
k+{1\over 3}({\lambda_p\over 4\pi})^3 k^2(3-{4\over
9}cos\theta)+0(\lambda_p^4)\eqno (64)$$ which is
$\theta$-dependent. For a pointlike, though regularized, potential
this is a surprising result. It hints at an interaction which is too
strong.\par
\null\hskip 1cm Next, fourth order. The result for the UV-divergent
contribution is $$f^{(4)}(k,\theta)=({\lambda_p\over 4\pi})^4
[{1\over
R^3}+{19\over 81}{k^2\over R}cos\theta -{1\over 3}{k^2\over
R}]+finite\null\hskip 2pt terms
\eqno (65)$$ There do not exist counterterms functions of
only R and eventually $\theta$ which make (65) finite. Perturbation
theory is not renormalizable. $\lambda_p$ has the wrong dimensions.\par
\null\hskip 1cm One could have thought of doing perturbations in
$\hat\lambda_p$ as defined in eq. (54). Then all perturbative
contributions
vanish. One could try intermediate choices, like $\lambda^\prime_p\equiv
R^{-1/2}\lambda_p(R)$. This only shifts the UV problem to higher orders.
There does not exist a non-trivial renormalizable perturbation theory.
There is no intermediate situation for the coupling between being too
weak and being too strong. A further insight into the UV structure
of this theory has been gained recently by studying the perturbative
$N>2$-body problem [33].\par
\null\hskip 1cm $\lambda\phi^4_{4+1}$ is a very similar QFT. It is
perturbatively non-renormalizable. The coupling has the wrong
dimensions. It is trivial, in a much more straightforward sense than
$\lambda\phi^4_{3+1}$, exactly as contact interactions in d=3 are
trivial in a more straightforward sense, due to the Thomas effect, than
in d=2.\par
\null\hskip 1cm For $d>3$ $\lambda\delta^{(d)}({\vec r})$ is
trivial for any $\lambda$, and perturbatively non-renormalizable. So is
$\lambda\phi^4_D$ for $D>4+1$.\par\null\vskip 1pt

\underbar {V. Conclusions}\par
\null\hskip 1cm If the WS theory follows the pattern of
singular QM no new physics is expected at 1 TeV on the grounds of the
existence of a Landau pole in the Higgs sector.
That weak interactions become strong only
reflects the inadequacy of truncated perturbation theory near a Landau
pole. That weak interactions are an effective
theory without reasonable continuum limit only follows from sticking to
(1). Weak interactions become as difficult at high energies as strong
interactions do at low energies. Both have ranges where perturbations
work, and ranges where they do not. Neither triviality nor Landau poles
anounce new physics, they are only a nuisance to theorists. So is
infrared slavery.\par
\null\hskip 1cm If the WS theory does not follow the pattern presented
here, we are still left with a, both exactly and perturbatively solvable
model, with a rich UV structure, and which gives a clear meaning to
Landau poles: they signal attractive perturbatively renormalizable
interactions, they are not
seen in physical magnitudes, they are only a problem when one truncates
perturbation theory which uses the wrong sign coupling and then takes
the continuum limit, or when one truncates perturbation theory which
uses the right sign coupling, and takes the low energy limit.\par
\null\hskip 1cm Our results also hint at one direction where theoretical
progress could be made in QFT when a Landau pole is nearby. The Landau
pole is seen in (2) or, equivalently, (42), but it is not seen in the
amplitude (43) because it has the coupling also in the denominator. One
needs beyond the running coupling a (even if partial) summation of the
very observable one is computing in perturbation theory, as e.g.
done with K-matrix methods and Pad\'e-approximants [34].\par
\null\hskip 1cm On the other hand the link between Landau poles and
triviality is not clear. Landau poles appear in the N=2 channel,
triviality for $N\rightarrow \infty$ (for d=2). Remember that in our
framework,
basically equivalent to a non-relativistic QFT but not the one
corresponding to the non-relativistic limit of the relativistic QFT we
are interested in, the Hilbert spaces of different N are not connected
to each other. In other words, is there a model of singular quantum
mechanics, perturbatively renormalizable, with Landau pole, and with a
non-trivial vacuum?\par
\null\hskip 1cm Our model also lacks non perturbative contributions, say
instantons, as perturbation theory reproduces the exact result. Another
relevant question thus is: is there a model of singular quantum
mechanics, perturbatively renormalizable, with Landau poles and
non-perturbative contributions? What do Landau poles mean then?\par

\null\hskip 1cm Let us finish with a final comment on singular QM: it is
so predictive, because it is very constrained by the combined
requirement of exact and perturbative renormalizability. So much so that
e.g. no model with two bound states (two Landau poles) is known.
Satisfying both requirements allows to go beyond what we can actually
acomplish for QFT. This is why it might add to our understanding of
QFT.\par

\null\vskip 2pt
ACKNOWLEDGEMENTS\par
\null\hskip 1cm I thank my colleagues Dom\`enec Espriu, Joaquim Gomis,
Jos\'e
Ignacio Latorre, Pedro Pascual, Josep M. Pons, Joan Soto and Josep Taron
for enjoyable
discussions and helpful comments. Financial support granted by CICYT,
AEN-93-0695 is acknowledged.\par
\vfill
\eject

\centerline {\bf REFERENCES}\par
\null\vskip 2pt
\settabs \+1aaaaa&\cr

\+1.&E. Fermi, Nuovo Cimento \underbar {11}, 1 (1934)\cr\par
\+2.&S. Weinberg, Phys. Rev. Lett. \underbar {19}, 1264 (1967)\cr\par
\+&A. Salam in "Elementary Particle Theory", ed. N. Svartholm (Almquist
and Wiksells,\cr\par
\+&Stockholm, 1969) p. 367\cr\par
\+&S.L. Glashow, J. Iliopoulos and L. Maiani, Phys. Rev. \underbar
{D2}, 1285 (1970)\cr\par
\+3.&L.D. Landau, "Fundamental problems" in "Collected papers of L.D.
Landau", eds. D. ter\cr\par
\+&Haar (Pergamon Press, Oxford, 1965)\cr\par
\+4.&M. Aizenman, Phys. Rev. Lett. \underbar {47}, 1 (1981)\cr\par
\+&J. Fr\" ohlich, Nuclear Physics \underbar {B200}, 281 (1982)\cr\par
\+5.&B.W. Lee, C. Quigg and H.B. Thacker, Phys. Rev. Lett. \underbar
{38}, 883 (1977)\cr\par
\+6.&D. Callaway, Phys. Reports \underbar {167}, 241 (1988)\cr\par
\+7.&M. L\" uscher and P. Weiss, Nuclear Physics
\underbar {B290}
[FS20], 25 (1987); Nuclear Physics\cr\par
\+&\underbar {B215} [FS21], 65 (1988); Physics
Letters \underbar {212B}, 472 (1988)\cr\par
\+8.&J. Gasser and N. Leutwyler, Phys. Reports \underbar {87C}, 77
(1982)\cr\par
\+9.&A. Manohar and H. Georgi, Nuclear Physics \underbar {B234}, 189
(1984)\cr\par
\+10.&S. Weinberg, Physica \underbar {A96}, 327 (1979)\cr\par
\+11.&K.G. Wilson, Rev. Mod. Physics \underbar {47}, 773 (1975)\cr\par
\+12.&H. Georgi, Annu. Rev. Nucl. Part. Sci. \underbar {43}, 209
(1993)\cr\par
\+13.&R. Dashen and H. Neuberger, Phys. Rev. Lett. \underbar {50}, 1897
(1983)\cr\par
\+14.&J. Glimm and A. Jaffe, "Quantum Physics" (Springer Verlag, New
York, 1981)\cr\par
\+15.&M. Consoli and P. Stevenson, Zeitsch. Physik \underbar {C63}, 427
(1994); "Resolution of the $\lambda\phi^4$\cr\par
\+&puzzle and a 2 TeV Higgs
boson" Rice preprint DE-FG05-92ER40717-5, July 1993\cr\par
\+16.&K. Huang, E. Manousakis and J. Polonyi, Phys. Rev. \underbar
{D35}, 3187 (1987)\cr\par
\+17.&S. Albeverio, F. Gesztesy, R. Hoegh-Krohn and H. Holder,
"Solvable
models in quantum\cr\par
\+&mechanics" (Springer Verlag, Berlin, 1988)\cr\par
\+18.&A.A. Belavin, A.M. Polyakov, A.S. Schwartz and Yu.S. Tyupkin,
Physics Letters \underbar {59B},\cr\par
\+&85 (1975)\cr\par
\+&G.'t Hooft, Phys. Rev. Lett. \underbar {37}, 8 (1976)\cr\par
\+&R. Jackiw and C. Rebbi, Phys. Rev. Lett. \underbar {37}, 172
(1976)\cr\par
\+19.&G.'t Hooft in "The Whys of Subnuclear Physics", Erice 1977, ed.
A. Zichichi (Plenum,\cr\par
\+&New York, 1979)\cr\par
\+20.&D.Z. Freedman, K. Johnson and J.I. Latorre, Nuclear Physics
\underbar {B371}, 353 (1992)\cr\par
\+&J.I. Latorre, C. Manuel and X. Vilas\'\i s-Cardona, Annals Physics
\underbar {231}, 149 (1994)\cr\par
\+21.&C. Thorn, Phys. Rev. \underbar {D19}, 639 (1979)\cr\par
\+22.&D.J. Gross and F. Wilczek, Phys. Rev. Lett. \underbar {30} 1343
(1973)\cr\par
\+&H.D. Politzer, Phys. Rev. Lett. \underbar {30}, 1346 (1973)\cr\par
\+23.&K. Symanzik, in Proc. Int. School of Physics "Enrico Fermi",
Varenna course XLV,\cr\par
\+&ed. R. Jost (Academic Press, New York, 1969)\cr\par
\+24.&M.A.B. B\'eg and R.C. Furlong, Phys. Rev. \underbar {D31}, 1370
(1985)\cr\par
\+25.&P. Gosdzinsky and R. Tarrach, Am. Journal Physics \underbar {59},
70 (1991)\cr\par
\+26.&R. Jackiw in "M.A.B. Beg memorial volume", eds. A. Ali and P.
Hoodbhoy (World\cr\par
\+&Scientific, Singapore, 1991)\cr\par
\+27.&A. Galindo and P. Pascual, "Quantum Mechanics" (Springer Verlag,
Berlin, 1991)\cr\par
\+28.&E.H. Lieb and W. Liniger, Phys. Rev. \underbar {130}, 1605
(1963)\cr\par
\+29.&C. Manuel and R. Tarrach, Physics Letters \underbar {B328}, 113
(1994)\cr\par
\+30.&G.F. Dell'Antonio, R. Figari and A. Teta, Ann. Inst. Henri
Poincar\'e \underbar {60}, 253 (1994)\cr\par
\+31.&L.H. Thomas, Phys. Rev. \underbar {47}, 903 (1935)\cr\par
\+32.&S.K. Adhikari, A. Delfino, T. Frederico, I.D. Goldman and L.
Tomio, Phys. Rev. \underbar {A37},\cr\par
\+&3666 (1988)\cr\par
\+33.&S.K. Adhikari, T. Frederico and I.D. Goldman, "Perturbative
renormalization in\cr\par
\+&quantum few-body problems", preprint Sao Paulo, Nov. 1994\cr\par
\+34.&D.A. Dicus and W.W. Repko, Phys. Rev. \underbar {D42}, 3660
(1990)\cr\par

\end